\documentclass[%
 reprint,
 amsmath,amssymb,
 aps,
]{revtex4-2}
\usepackage{xcolor}
\usepackage{graphicx}
\usepackage{dcolumn}
\usepackage{bm}
\usepackage{multirow}
\usepackage{hyperref} 

\begin{document}

\preprint{APS/123-QED}

\title{Holographic quark masses and radiative decays of heavy vector  mesons}

\author{Saulo Diles}
\email{smdiles@ufpa.br}
\affiliation{Campus Salin\'opolis,\\ Universidade Federal do Par\'a,\\
68721-000, Salin\'opolis, Par\'a, Brazil}
\affiliation{Unidade Acadêmica de Física,\\ Univ. Federal de Campina Grande, R. Aprígio Veloso, 58429-900 - Campina Grande}

\author{Miguel Angel Martin Contreras}
\email{miguelangel.martin@usc.edu.cn}
\affiliation{
 School of Nuclear Science  and Technology\\
 University of South China\\
 Hengyang, China\\
 No 28, West Changsheng Road, Hengyang City, Hunan Province, China.
}
\affiliation{Key Laboratory of Advanced Nuclear Energy Design and Safety,\\ Ministry of Education, 28 Changsheng West Road, Hengyang 421001, China.}

\author{Alfredo Vega}%
 \email{alfredo.vega@uv.cl}
\affiliation{%
 Instituto de F\'isica y Astronom\'ia, \\
 Universidad de Valpara\'iso,\\
 A. Gran Breta\~na 1111, Valpara\'iso, Chile
}

\date{\today}   

\begin{abstract}
Holographic models of QCD provide the spectrum of heavy vector meson masses and electromagnetic decay constants through bulk computations of the current-current correlation function. Conversely, the phenomenology of heavy vector mesons is articulated by the constituent heavy quark model utilizing a non-relativistic approximation. By applying the Segre formula from non-relativistic quantum mechanics, we derive new observables from holography: the constituent quark mass, the three-photon decay width, the effective fine structure constant of the strong interaction, and the mixed one-photon and two-gluon decay width. We also derive the three-gluon decay width,   the three-photon decay width, and the mixed one-photon and two-gluon decay width for the radially excited states of heavy quarkonia and compare them with available experimental data. The present results reveal a new paradigm of meson spectroscopy in AdS/QCD.
\end{abstract}
\maketitle


\section{\label{sec:level1} Introduction}

The Anti-de Sitter/Conformal Field Theory (AdS/CFT) correspondence establishes a mapping between expectation values of the product of operators in a quantum conformal field theory and dual (semi-classical) fields that reside in a higher-dimensional Anti-de Sitter space \cite{Maldacena:1997re, Gubser:1998bc}. Specifically, in the context of vector mesons, the expectation value of the current-current correlation function, analyzed in momentum space, encapsulates comprehensive information regarding the spectrum of vector mesons, including their masses and electromagnetic decay constants associated with an infinite series of radial excitations of the quark-antiquark ($q \bar{q}$) confined system. The fundamental concept underpinning AdS/QCD models involves the introduction of a bulk scalar field (dilaton) that effectively generates a mass scale within the bulk geometry. This mass scale breaks the conformal invariance of the associated dual gauge theory, which plays a crucial role in inducing confinement \cite{Karch:2006pv}.

In the holographic framework, a meson is produced by inserting an operator with the appropriate conformal dimension, and it does not reference a constituent quark structure. This represents a coherent methodology for discussing hadron properties \cite{Fritzsch:1972jv}. In this context, holographic hadrons are not required to exhibit an internal structure. However, it is well established that hadrons possess an internal structure characterized by valence quarks, which delineate their physical properties \cite{Gell-Mann:1964ewy, Lucha:1991vn}. The bottom-up AdS/QCD approach has been widely used in the computation of electromagnetic decay constants \cite{Erlich:2005qh, Braga:2015jca}. It is observed that the experimental value of the decay constant $f$ is indirect; we assume a constituent quark model with massive quarks, with electromagnetic charge $Q_q$, which subsequently links the direct measurement of the electron-positron decay width $\Gamma_{V\to e^+e^-}$ with the decay constant \cite{VanRoyen:1967nq, Hwang:1997ie}:

\begin{equation}\label{decayf}
    f_n^2 = \frac{3\,M_n}{4\,\pi\,\alpha_{\textrm{em}}^2\,Q_q}\,\Gamma(q\bar{q}\to e^+e^-).
\end{equation}

The reference to a constituent quark model manifests in the electric quark charge. Indeed, the physical process of meson vacuum decay directly results from the flavorless structure of its constituents. When we seriously consider the existence of a constituent quark model within the AdS/QCD paradigm, we gain new insights into meson physics.   

The AdS/QCD model defines a hadron spectrum of masses and decay constants $\{M_n,f_n\}$, which we associate with a confined state of quarks. We focus on heavy vector mesons, whose decay width for annihilation is modeled using non-relativistic quantum mechanics \cite{Caswell:1985ui, Bodwin:1994jh}. The constituent quark model includes many observables, such as the heavy quark masses, electric quark charges, the electromagnetic fine structure constant, strong force fine structure, and, for the vector mesons, four annihilation decay channels: electromagnetic, three-photons, three-gluons, and one-photon-two-gluons.

This manuscript is organized as follows: in Section
 \ref{hadron}, we discussed how hadrons are described in bottom-up models. In Section \ref{Segre}, we introduced the Fermi-Segre formula and calculated the quark constituent mass from the hadronic spectrum. Section \ref{decays} discussed heavy vector meson decays and the strong coupling constant using the holographic Segre formula. In Section \ref{highly}, we extend our calculations to the case of highly radial excited states, allowing us to include relativistic corrections to the holographic Segre formula. Finally, we deliver our main conclusions in \ref{concl}. 

\section{Hadrons in AdS/QCD}\label{hadron}
In the AdS/CFT correspondence context, hadrons are described as non-perturbative boundary operators corresponding to bulk fields, and for a given hadronic boundary operator, expressed as $\mathcal{O}=f(q,\bar{q}, G_{\mu\nu}, D_\mu)$, the scaling dimension $\text{dim}\,\mathcal{O}=\Delta+L+\gamma$ defines the hadronic identity in bulk AdS via the bulk mass $M_5$ \cite{Aharony:1999ti}. Here, $\Delta$ is the energy dimension of the constituent operator, $L$ is the angular momentum, and $\gamma$ represents the anomalous dimension. This relationship results from the field/operator duality, which dictates how the bulk fields reflect the boundary data \cite{Polchinski:2002jw}. Therefore, a dual bulk field to a given hadron state has a near-conformal boundary behavior given by the dimension of the operator
$\text{dim}\,\mathcal{O}$. This duality is reflected in the bulk field mass $M_5=M_5(\Delta, L,\gamma)$. Thus, this relation defines the \emph{hadronic identity}. However, in this definition, we only have information about the quantity of constituents ($\Delta$) but not about the inner configuration. Thus, in the holographic context, \emph{a hadron is a bag of constituents with no information about hadronic inner structure}, characterized by the bulk field mass $M_5$.   

This paper will focus on vector mesons and consider vector bulk fields. Nevertheless, this analysis can be extended to hadronic species such as scalar mesons or baryons. 

The starting point is defining the geometric background given by the Poincarè patch:

\begin{equation}
    dS^2=\frac{R^2}{z^2}\left[dz^2+\eta_{\mu\nu}\,dx^\mu\,dx^\nu\right],
\end{equation}

\noindent where $R$ is the AdS radius, $z$ is the holographic coordinate, and $\eta_{\mu\nu}$ is the $4$-dimensional Minkowski metric with $(-,+,+,+)$ signature. We will follow the \emph{softwall-like} approach in this geometric context. In this framework,  Hadrons are introduced using bulk fields defined by the bulk action:

\begin{multline}\label{action-hadron}
    I_
\text{Hadron}=-\frac{1}{2\,g_5^2}\,\int{d^{5}x\,\sqrt{-g}\,e^{-\Phi(z)}}\left[F_{MN}\,F^{MN}\right.\\
\left.+M_5^2\,A_M\,A^M\right],
\end{multline}

\noindent where $g_5$ fixes the decay constants, $\Phi(z)$ is a \emph{static} dilaton field, $A_M$ is the bulk vector field dual to vector hadrons at the conformal boundary, $F_{MN}=2\partial_{(M}\, A_{N)}$ is the associated field strength. The \emph{Latin} indices label $5$-dimensional directions in the AdS space.

Confinement, understood as the emergence of bound states of quarks in this model, is achieved by the dilaton field. In pure AdS space, bulk modes are not normalizable due to field/operator duality: states localized at the boundary become delocalized and non-normalizable, implying no confinement \emph{ab initio}. The dilaton field can resolve this issue by introducing an energy scale determining the hadronic masses \cite{Karch:2006pv}. This idea is precisely the core of the bottom-up holographic QCD. The \emph{bottom-up magic} occurs in the so-called confining holographic potential, defined from the bulk equations of motion.   

How do we obtain the confining holographic potential? We start from the equation of motion for the bulk fields derived from the action \eqref{action-hadron}:

\begin{equation}
 \frac{1}{\sqrt{-g}}\,\partial_m\left[\sqrt{-g}\,g^{mr}\,g^{np}\,F_{rp}\right]-M_{5}^2\,g^{mn}\,A_m=0.
\end{equation}

\noindent where the bulk mass for a bulk vector field is defined as \cite{Aharony:1999ti}:

\begin{equation}
    M_5^2\,R^2=\left(\Delta+L+\gamma-1\right)\left(\Delta+L+\gamma-3\right).
\end{equation}

Imposing the \emph{axial gauge} $A_z=0$ is customary, implying that at the boundary we will have $\partial_\mu\, A^\mu=0$.
Since we will consider vector mesons ($\Delta=3$) in $s$-wave, with no other additions from anomalous dimensions, we can fix $L=\gamma=0$. Therefore, the bulk mass for vector mesons reduces to zero: $M_5^2\, R^2=0$. After all of these considerations, we obtain the expression: 

\begin{equation}\label{eqn-mo}
z^3\,\partial_z\left[z^{-1}\,e^{-\Phi}\,\partial_z\,A_\mu\right]+z^2\,e^{-\Phi}\,\Box\,A_\mu=0. 
\end{equation}

We Fourier transform the bulk field and decompose it as $A_\mu(z,q)=\tilde{A}_\mu\,\psi(z,q)$, where $\tilde{A}_\mu(q)$ is the source operator at the boundary and $\psi(z)$ defines the \emph{bulk mode} dual to vector mesons, the eqns. \eqref{eqn-mo} reduce to \emph{Sturm-Liouville form}: 

\begin{equation}\label{eqn:sturm}
    \partial_z\left[e^{-B(z)}\,\partial_z\,\psi(z,q)\right]+M_n^2\,e^{-B(z)}\,\psi(z,q)=0.  
\end{equation}

\noindent where we have set \emph{on-shell condition} as $-q^2=M_n^2$, and have introduced $B(z)=\Phi(z)-\log\,\left(R/z\right)$. The next step in the bottom-up recipe is applying the \emph{Boguliubov transformation} $\psi(z)=e^{\Phi(z)/2}u(z)$ that converts eqn. \eqref{eqn:sturm} into a \emph{Schr\"odinger-like} one, emerging the holographic confining potential $V(z)$ in the process:

\begin{equation}\label{Eq-Schr}
    -u''(z)+V(z)\,u(z)=M_n^2\,u(z),
\end{equation}

\noindent with 

\begin{equation}\label{eq:holo-pot}
    V(z)=\frac{3}{4\,z^2}+\frac{1}{2\,z}\Phi'(z)
   +\frac{1}{4}\Phi'(z)^2-\frac{1}{2}\Phi''(z).
\end{equation}

Let us make some comments on the potential. Notice that the first term comes from the AdS warp factor entirely, which is nonconfining.      If we turn off the dilaton, the holographic potential does not generate bounded states unless a barrier is added, as in the quantum infinite well. This is the case of the \emph{hardwall model} \cite{Polchinski:2002jw, Boschi-Filho:2002glt},   where a spatial hard cutoff realizes confinement. The cutoff locus defines the mass-energy scale. The eigenvalue spectrum obtained from the holographic potential defines the set of radial hadronic Regge trajectories $M_n^2$, which, depending on the constituent masses, could be linear or not \cite{MartinContreras:2020cyg}. The high-$z$ behavior of the potential controls the Regge trajectory structure. A WKB analysis supports this claim \cite{MartinContreras:2025mpj}.

The last remark deals with bulk eigenmodes. If we change the boundary conditions in the Sturm-Liouville equation from $\psi(z\to0)=\mathcal{C}\,z^{\Delta-1}$ to $\psi(z\to0)=1$, the bulk field becomes the \emph{bulk-to-boundary propagator} $\mathcal{V}(z,q)$. If we now evaluate the bulk action \eqref{action-hadron} at the conformal boundary ($z\to0$) for $\mathcal{V}(z,q)$, we can compute the holographic $2$-point function by applying the GPK-Witten relation \cite{Aharony:1999ti}, 

\begin{equation}
\langle 0\left|\mathcal{T}\,J_\mu(q)\,J_\nu(0)\right|0\rangle=-\frac{\delta^2\,I_{\text{Bndry}}^{On-shell}}{\delta\,A^{\mu,0}(q)\,\delta\,A^{\nu,0}(0)}    
\end{equation}
which is dual to the large $N_c$ QCD vector current 2-point function \cite{Erlich:2005qh}: 

\begin{eqnarray}
 \langle 0\left|\mathcal{T}\,J_\mu(q)\,J_\nu(0)\right|0\rangle&=&\left(q^2\,\eta_{\mu\nu}-q_\mu\,q_\nu\right)\,\Pi_V\left(-q^2\right),\\
 \Pi_V\left(-q^2\right)&=&\frac{1}{g_5^2\,\left(-q^2\right)}e^{-B(z)}\,\left.\frac{\partial\mathcal{V}(z,q)}{\partial z}\right|_{z\to0}.
\end{eqnarray}

The last term, the \emph{polarization operator}, allows us to find the value of $g_5$ compared to the large $N_C$ QCD 2-point function at the large $q^2$ limit \cite{Shifman:1978bx}, which produces $g_5^2=12\,\pi^2/N_c$.  

At the holographic level, the holographic expression for the polarization factor has information about the holographic vector meson duals. Recall that the bulk-to-boundary propagator can be \emph{spectrally}
decomposed in terms of the bulk normalizable modes $\psi_n(z,q)$ as poles and residues as

\begin{equation}
     \Pi_V\left(-q^2\right)=\sum_{n=1}^\infty\frac{f_n^2}{p^2-m_n^2+i\epsilon}.
\end{equation}

\noindent corresponding to the vector meson mass spectrum, and the \emph{holographic decay constants} $f_n$ defined as \cite{MartinContreras:2019kah}:

\begin{equation}\label{eqn:decay-cons}
    f_n^2=\frac{1}{g_5^2\,M_n^2}\,\lim_{\varepsilon\to0}\,e^{-2\,B(\varepsilon)}\left|\psi_n(\varepsilon,q)\right|^2. 
\end{equation}

Therefore, the decay constants depend on the bulk normalizable modes and the dilaton profile near the conformal boundary, as originally suggested in \cite{Braga:2015jca}. This characteristic plays an important role in studying the phenomenology of heavy vector mesons since properly modeling decay constants allows us to uncover thermal \cite{MartinContreras:2021bis} and finite-density \cite{Braga:2017bml} effects. 

Another important motivation for approaching this definition of decay constants is the connection with the hadronic wave function at the origin through the so-called \emph{Van Royen-Weisskopf formula} \cite{VanRoyen:1967nq}. We will explore this connection in the next section. 

\section{Segre formula}\label{Segre}
In the context of non-relativistic quantum mechanical modeling of heavy quarkonium, we identify a relationship, referred to as the Segre formula, between the radial wave function at the origin for a radial excitation of the $s$-wave state $\Psi_n(0)$ and the dependence of the binding energy on the excitation level $E_n$. This relationship establishes that \cite{Quigg:1977wn}:

\begin{equation}
    |\Psi_n(0)|^2=\frac{(2\mu)^{\frac{3}{2}}}{4\pi^2}E_n^{\frac{1}{2}}\frac{d}{dn}E_n.
\end{equation}

The formula complements the results of AdS/QCD. In fact, AdS/QCD models compute the mass and decay constant spectrum encoded in the current-current correlator. However, this approach cannot explore hadronic binding energies or inner structure configurations since AdS/QCD does not provide quark mass information \emph{a priori}.  Recall that, at the holographic level, hadrons are characterized only by their constituent content via the bulk mass. In top-down approaches \cite{Erdmenger:2007cm}, the quark mass emerges as an energy scale when geometrical perturbations are included. However, it does not bring any outcomes about hadronic inner structure properties.  
 
From non-relativistic QCD models, the experimental data regarding meson decay is used to extract quark masses from the decay constant ratio between consecutive radial excitations \cite{Lucha:1991vn}. We can follow the same approach in bottom-up AdS/QCD since decay constants can be computed using relations such as eqn. \eqref{eqn:decay-cons} to extract \emph{heavy} quark masses.

From a QCD perspective, the electromagnetic (EM) vector meson decay width is expressed according to eqn. \eqref{decayf}. At the lowest order in perturbation theory, the EM decay width is related to the spatial wave function at the origin as:

\begin{equation}
    \Gamma_{V_n\to e^+e^-} = \frac{\pi\alpha_{\textrm{em}}^2\,Q_q^2}{M_n^2}\lim_{r\to0}\left|\Psi_n(\vec{r})\right|^2.
\end{equation}

Therefore, according to \emph{Van Royen-Weisskopf formula}, the mesonic decay constants can be expressed in terms of the spatial wave function at the origin as 

\begin{equation}\label{decayspace}
    f_n^2 = \frac{12\,\pi \,Q_q }{M_n}\,\left|\Psi_n(0)\right|^2,
\end{equation}

\noindent which is independent of the electromagnetic fine structure constant. Moreover, the annihilation rates of mesons are always related to the probability that the quark will reach the antiquark, i.e., the norm of the wave function at the origin. Since the electric charge and masses of the constituent quarks are identical for each family of vector mesons, the ratio of wave functions at the origin for different excited states is completely determined by spectroscopy. 

\begin{equation}\label{ratio}
    \frac{|\Psi_{n_1}(0)|^2}{|\Psi_{n_2}(0)|^2}=\frac{M_{n_1}f_{n_1}^2}{M_{n_2}f_{n_2}^2}.
\end{equation}

The equation above tells us that the holographic mass and decay constant spectra fix the ratio of the wave function at the origin at the boundary. On the other hand, to use the Segre-Fermi formula holographically for heavy vector mesons, we define the meson mass (obtained by solving the Schrödinger-like equation) in terms of the binding energy as

\begin{equation}
    M_n = 2m_q+E_n.
\end{equation}

Thus, collecting all of the previous results and definitions, we can write an expression for the quark masses in terms of the holographic spectra of vector meson masses and decay constants:

\begin{equation}\label{eq:holo-segre}
  \frac{|\Psi_{n_1}(0)|^2}{|\Psi_{n_2}(0)|^2}=\frac{M_{n_1}f_{n_1}^2}{M_{n_2}f_{n_2}^2}=\frac{\left(M_{n_1}-2\,m_q\right)^{1/2}}{\left(M_{n_2}-2\,m_q\right)^{1/2}} \frac{\left.\frac{d\,M_{n}}{d\,n}\right|_{n=n_1}}{\left.\frac{d\,M_{n}}{d\,n}\right|_{n=n_2}}. 
\end{equation}

The relativistic effects are implemented in this context by a scaling factor of the wave function at the origin \cite{Durand:1983bg}:

\begin{equation}\label{relcorrect}
    \left|\Psi^{\text{rel.}}_n(0)\right|^2 = \left(1+\frac{E_n}{m_q}\right) \left(1+\frac{E_n}{4m_q}\right)^{\frac{1}{2}} |\Psi_n(0)|^2.
\end{equation}

In summary, once the dilaton is fixed, we use equation \eqref{Eq-Schr} to compute the radial Regge trajectories from the holographic confining potential \eqref{eq:holo-pot} and the decay constant spectrum from the bulk modes, following the expression \eqref{eqn:decay-cons}. Next, we parametrize the Regge trajectories and use the holographic Segre-Formula \eqref{eq:holo-segre} to compute the constituent quark masses. Let us see how this works. 

In heavy meson phenomenology, Regge trajectories are expected to deviate from linearity due to the heavy quark mass \cite{Chen:2018bbr, MartinContreras:2019kah}. Therefore, the radial Regge trajectories can be parametrized as

\begin{equation}
    M_n^2=a(n+b)^\nu,
\end{equation}

\noindent where $a$ is the energy scale that fixes the mass, proportional to the string tension, $b$ is the intercept, and $\nu$ is the \emph{linearity deviation}. Thus, in the context of the Braga non-quadratic dilaton \cite{MartinContreras:2021bis}:

\begin{equation}
    \Phi(z)=\left(\kappa\,z\right)^{2-\alpha}+M\,z+\tanh\left[\frac{1}{M\,z}+\frac{\kappa}{\sqrt{\Gamma}}\right]
\end{equation}

\noindent we can solve for the charmonium and bottomonium first two states\footnote{These softwall-like models are \emph{flavor-dependent}, meaning that for each meson family, we need a set of parameters $\{\kappa,\,\alpha,\,M,\,\,\text{and}\,\,\Gamma\}$.}, yielding the constituent quark masses for the charm and bottom quarks as

\begin{equation}
\begin{aligned}
        m_c &= 1.97\pm 0.28~\text{GeV},\\
        m_b&=5.21\pm 0.36~\text{GeV}. 
\end{aligned}
   \end{equation}
   
Notice that the quark masses obtained are \emph{constituent masses}, different from the \emph{current ones}. In constituent quark models, it is customary to assume that constituent masses could be fixed as half of the $1S$ state mass, see \cite{Scadron:2006dy, Deng:2016stx}. However, this is not a universal rule in these models, and we observe a good agreement between our results and those of other constituent quark models, such as \cite{Celmaster:1977vh, Stanley:1980zm, Segovia:2016xqb}.

\section{Heavy vector meson decays}\label{decays}
\subsection{Three gluons decay and the  fine structure constant of the strong interaction}

The mesonic decay of mesons into gluons is an interesting tool for probing QCD dynamics. For vector mesons, there is a relation between the wave function at the origin and the total meson decay width into three gluons. It is given by \cite{Kwong:1987ak, Lucha:1991vn}:

\begin{equation}\label{totalhadron}
    \Gamma_{V_n\to\textrm{ggg}}= \frac{40(\pi^2-9)}{81m_q^2}\,\alpha_{\textrm{s}}^3\left|\Psi_n(0)\right|^2.
\end{equation}

In our holographic context, the bottom-up model provides the spectral properties of mesons, i.e., holography provides $\{M_n,f_n\}$ for a given meson family. Indeed, as demonstrated in eqn. \eqref{decayspace}, we have expressed the wave function at the origin through its spectral properties: 

\begin{equation}
    \left|\Psi_n(0)\right|^2=\frac{f_n^2\,M_n}{12\pi\,Q_q}.
\end{equation}

Here we shall adopt the quark electromagnetic charges as the known fractions of the elementary charge,  i.e, $Q_c=\frac{2}{3}$  and $Q_b=\frac{1}{3}$, and the fine electromagnetic structure is well approximated as $\alpha_{\textrm{em}}\simeq\frac{1}{137}$. We use the experimental measurement of the three-gluon decay width to establish the effective strong coupling constant, which we use to compute all the decay widths at leading order in $\alpha_s$. In this sense, we can use the PDG information \cite{ParticleDataGroup:2024cfk} on $J/\psi$ and $\Upsilon$ mesons to define $\alpha_s(c)$ and $\alpha_s(b)$ from holographic data. Thus, we find $\alpha_s(q)$ from the expression for the three-gluon decay as

\begin{equation}
    \Gamma_{V_1\to ggg} =\frac{10(\pi^2-9)\,f_1^2\,M_1}{243\,\pi\,m_q^2\,Q_q}\alpha_s(q)^3.
\end{equation}

Using the results from the Braga non-quadratic model, as well as the quark masses obtained from the Segre formula, we obtain:

\begin{equation}\label{leadingalpha}
 \alpha_s(c) = 0.293\pm0.008,~~\alpha_s(b)=0.192\pm0.009.
\end{equation}

Notice that these values are close to the calculations of \cite{Kwong:1987ak}, which reinforces the consistency of the holographic modeling of quark masses. We can extrapolate this result to the next order in $\alpha_s$ corrections. 

\subsection{The first-order corrections on $\alpha_s$}
The three-gluon decay for vector heavy quarkonium, eqn.  \eqref{totalhadron} is the dominant channel in the total decay. Therefore, by evaluating this process, we can extract a value of the strong coupling constant for charmonium or bottomonium.  We will define a value for $\alpha_{\textrm{strong}}$ to properly implement first-order corrections to different channels in the vector meson decay. 

In the previous subsection, we calculated the strong coupling constants $\alpha_s$ at \emph{zero} order from the total hadron decay width. Despite that this expression is cubic in $\alpha_s$, it receives first-order corrections \cite{Barbieri:1980yp, Kwong:1987ak}. Since our holographic model provides meson properties from a two-point correlator function, which encodes the spectrum of masses and decay constants, we express decay widths in terms of these spectral properties, rather than the radial wave function. 

Let us first consider the corrections to the electromagnetic decay constants induced by considering $\alpha_s$:

\begin{equation}
     \Gamma_{V_n\to e^+e^-} = \frac{\pi\alpha_{\textrm{em}}^2\,Q_q^2}{M_n^2}|\Psi_n(0)|^2\left(1-\frac{\alpha_s}{\pi}\frac{16}{3}\right). 
\end{equation}

The decay constants $f_n$ should not be sensitive to corrections of $\alpha_s$, since they are defined directly from experimental measurements of the $e^+e^-$ decay width. Therefore, we must properly compute wave function corrections at the origin, which are $\alpha_s$ dependent. At first order, we have the following:

\begin{equation}
    \left|\Psi_n(0)\right|^2=\frac{f_n^2\,M_n}{12\pi\,Q_q}\left(1+\frac{\alpha_s}{\pi}\frac{16}{3} \right).
\end{equation}

Since the $\alpha_s$ corrections on $|\Psi_n(0)|^2$ do not depend on $n$, there is no change in eqn. \eqref{ratio}. Therefore,  the constituent quark masses are \emph{invariant} under higher-order $\alpha_s$ corrections.

We express the three-gluon decay width regarding the holographic meson spectrum, keeping first-order corrections in $\alpha_s$, and compare with the experimental measurement. Thus, defining the holographic value of $\alpha_s$ at first-order, we have:

\begin{equation}\label{3gluon}
\begin{aligned}
    \Gamma_{V_1\to ggg} =&\frac{10(\pi^2-9)\,f_1^2\,M_1}{243\,\pi\,m_q^2\,Q_q}\\
    \times \alpha_s^3&\left[1+\frac{\alpha_s}{\pi}\left(\frac{16}{3}-\lambda_q \right)\right],    
\end{aligned}
\end{equation}

\noindent where $\lambda_q$ encodes a characteristic energy associated with each meson family and depends only on the number of lighter quark flavors. For charmonium $\lambda_c=3.7$, and for bottomonium $\lambda_b=4.9$.  Next, we obtain the first-order $\alpha_s$ for the charm and bottom quark families as:

\begin{equation}
  \alpha^1_s(c) = 0.280\pm0.007,~~\alpha^1_s(b)=0.191\pm0.009.
\end{equation}

From now on,  we perform our calculations considering first-order corrections in $\alpha_s$, allowing us to omit the superscript $^1$ everywhere. We observe that our holographic calculation using the experimental measurement of the three-gluon decay width agrees with the theoretical approach used in \cite{Godfrey:2015dia}, where $\alpha_s(b)$ is defined from the quark mass scale and a renormalization scheme. We note that for the charm quark, the change due to first-order corrections on its definition is about $5\%$, while for the bottom quark it is less than $1\%$. 

We obtain the three-gluon decay width for the excited states using eqn.\eqref{3gluon}, which we summarize in Table \ref{tab:one}. Our result for $\Psi'$ is precise (less than $1\%$). On the other hand, for $\Upsilon'$ and $\Upsilon''$, our results are overestimated. 

\begin{center}
\begin{table*}[t] 
    \begin{tabular}{|c|c|c|c|c|c|c|c|}
    \hline
    \hline
    \multicolumn{8}{c}{\textbf{Summary of Data}}\\
    \hline 
    \hline
    \textbf{Quarkonium} & \textbf{State} & \multicolumn{2}{|c|}{$\Gamma_{V\to ggg}$} & \multicolumn{2}{|c|}{$\Gamma_{V\to \gamma \gamma \gamma}$}& \multicolumn{2}{|c|}{$\Gamma_{V\to \gamma gg}$}\\
    \hline
    \hline
    \multirow{4}{2em}{\textbf{$c\bar{c}$}} & $n^{2S+1}\,L_J$ & \textbf{Exp. KeV} & \textbf{Theo. KeV } & \textbf{Exp. eV} & \textbf{Theo. eV } & \textbf{Exp. KeV} & \textbf{Theo. KeV }\\
    \cline{2-2} \cline{3-3} \cline{4-4} \cline{5-5}\cline{6-6}\cline{7-7}\cline{8-8}
     & $1^3\,S_1$ & $59.3\pm1.43^*$ & $59.3\pm1.43^*$ & $1.07\pm 0.20$& $0.96\pm0.09$  &$8.15\pm 1.31$ & $3.79\pm0.36$ \\
     & $2^3\,S_1$ &$30.3\pm4.9$  & $30.22\pm5.88$ & --& $0.49\pm0.09$ & $2.94\pm 0.84$& $1.93\pm0.35$\\
     & $3^3\,S_1$ & -- & $18.36\pm3.57$ & -- & $0.30\pm0.05$& --& $1.17\pm0.21$\\
     & $4^3\,S_1$ & -- & $13.09\pm1.59$ &-- & $0.21\pm0.02$& --& $0.83\pm0.08$ \\
    \hline
    \hline
    \multirow{4}{2em}{\textbf{$b\bar{b}$}} & $n^{2S+1}\,L_J$ & \textbf{Exp. KeV} & \textbf{Theo. KeV } & \textbf{Exp. eV} & \textbf{Theo. eV } & \textbf{Exp.} \textbf{KeV} & \textbf{Theo. KeV}\\
    \cline{2-2} \cline{3-3} \cline{4-4} \cline{5-5}\cline{6-6}\cline{7-7}\cline{8-8}
     & $1^3\,S_1$ & $44.13\pm1.09^*$ & $44.13\pm1.09^*$ & -- & $0.0627\pm0.0095$ & $1.18\pm0.32$ & $1.15\pm0.19$\\
     & $2^3\,S_1$ & $18.80\pm1.59$ & $29.46\pm11.12$&--& $0.0419\pm0.0147$ & $0.60\pm0.10$ & $0.77\pm0.28$\\
     & $3^3\,S_1$ & $7.25\pm0.84$ & $6.21\pm1.33$&-- &$0.0088\pm0.0014$ & $0.20\pm 0.04$& $0.16\pm0.03$\\
     & $4^3\,S_1$ & -- & $4.14\pm0.91$ & -- & $0.0059\pm0.0010$ & -- & $0.11\pm0.02$\\
     \hline
     \hline
    \end{tabular}
    \caption{Summary of the results for the three-gluon, three-photon and $g\gamma\gamma$ decays for the heavy quarkonium vector ($1^3S_n$) trajectories. Experimental results are read from PDG \cite{ParticleDataGroup:2024cfk}. The asterisk (*) indicates that we used this data for fitting the corresponding decay width. }
    \label{tab:one}
\end{table*}    
\end{center} 

\subsection{Three-photon decay}

Another important implication of defining quark masses from the Segre formula is the calculation of other decay widths, usually written in terms of the normalized wave function at the origin. In this context, the three-photon decay emerges as a promising probe. This decay is a pure QED process at leading order, mediated by a quark-loop diagram. It is possible to analyze high-order (loop) corrections and quarkonium structure (by testing $|\Psi(0)|^2$), relevant for NRQCD potential models \cite{Brambilla:2022ayc}. Three-photon decay of $J/\Psi$ was first detected in 2008 \cite{CLEO:2008qfy}, updated in 2013 \cite{BESIII:2012lxx}, and calculated using lattice QCD in 2020 \cite{Meng:2019lkt}. The theoretical discussion first appears in the context of positronium atoms \cite{ore1949three}, which can be thought of as the leptonic analog of the mesons. 

The vector meson three-photon decay is calculated in ref. \cite{Godfrey:2015dia}, and reads: 

\begin{equation}
\begin{aligned}
    \Gamma_{V_n\to 3\gamma} =& \frac{16(\pi^2-9)}{3m_q^2}\,Q_q^6\,\alpha_{\textrm{em}}^3\left|\Psi_n(0)\right|^2\times \\&\left[1-\left(\frac{12.6}{\pi }-\frac{16}{3 \pi }\right) \alpha_{s}(q)\right].
\end{aligned}
\end{equation}

Subsequently, we represent the three-photon decay without explicitly referencing the radial wave function:

\begin{equation}
\begin{aligned}
    \Gamma_{V_n\to 3\gamma} =& \frac{4 (\pi^2-9)\,Q_q^5\,\alpha_{\text{em}}^3\,f_n^2\,M_n}{9 \,m_q^2} \\&\left[1-\left(\frac{12.6}{\pi }-\frac{16}{3 \pi }\right) \alpha_{s}(q)\right].
    \end{aligned}
\end{equation}

For Charmonium, at first-order in $\alpha_s$, we find that $\Gamma_{J/\Psi\to3\gamma} = 0.96\pm0.09~\text{eV},$ and for bottomonium $\Gamma_{\Upsilon(1S)\to3\gamma} = 0.0627\pm0.0095~\text{eV}$. The results for excited states are summarized in Table \ref{tab:one}. The result for $J/\Psi$ is underestimated. Nevertheless, it exhibits the same order of magnitude as its experimental measure, which is $\Gamma_{J/\Psi\to 3\gamma}=1.07\pm 0.20~\text{eV}$ \cite{BESIII:2012lxx}. It is important to note that no experimental data are available for charmonium and bottomonium excited states. We also note that our result differs in one order of magnitude from the theoretical model exposed in Ref. \cite{Segovia:2016xqb}, where it is found $\Gamma_{\Upsilon(1S)\to3\gamma}\sim 0.003 \text{eV}$.

\subsection{The mixed strong and electromagnetic decay}

Once we have holographically computed quark masses and effective strong couplings for the heavy quark systems, we can use these results and compute another heavy vector meson decay width: the one photon and two gluon decay $q\bar{q}\to \gamma gg$. We follow the discussion given in Refs. 
\cite{Soni:2017wvy, Kher:2018wtv}  to obtain the appropriate expression for this decay channel:

\begin{multline}
    \Gamma_{V_n\to\gamma gg}=\frac{ 8 \left(\pi ^2-9\right) Q_q\, \alpha_{s}^2(q)\, \alpha_{\text{em}}\,M_n\, f_n^2}{27\pi\, m_q^2}\times\\ \left[1-\left(\frac{\sigma_q}{\pi }-\frac{16}{3 \pi }\right) \alpha_{s}(q)\right],
\end{multline}

\noindent where $\sigma_q$ parameterize the first-order corrections in $\alpha_s$: $\sigma_c=6.7,~\sigma_b=7.4$. 

Our results for both ground and excited states, along with the available experimental data, are summarized in Table \ref{tab:one}. The $J/\Psi$ width is underestimated, while the $\Psi'$  width belongs to the experimental data confidence interval. For bottomonium, all decay widths are overestimated. However, they have the same order of magnitude, indicating that our results are consistent.  Notice that our results accurately represent the correct phenomenological behavior, as they decrease with increasing excitation level. Other potential models provide different theoretical predictions, see \cite{Godfrey:2015dia, Segovia:2016xqb, Kher:2018wtv, Garg:2023mst}.

\section{Highly excited states}\label{highly}
The notion of decay constants in a constituent quark model is defined by the value of the corresponding Schrödinger wave function at the origin,  $\Psi_n(0)$. In \cite{Durand:1983bg}, this picture is discussed in detail, obtaining the Segre formula in its original non-relativistic version and its relativistic counterpart, i.e., $\Psi^{\small{\textrm{rel.}}}_n(0)$. For the non-relativistic Segre formula, we have:

\begin{equation}
    |\Psi_n(0)|^2 = \frac{m_q^2}{4\pi^2}v_n\frac{d E_n}{dn},
\end{equation}

\noindent where $v_n=\sqrt{\frac{E_n}{m_q}}$ is the velocity scale in the bounded state. It leads to

\begin{equation}
       \left|\Psi_n(0)\right|^2\propto\frac{d}{dn}\, E_n^{\frac{3}{2}}.
\end{equation}

On the other hand, sum rules discussions provide independent information about the dependence of the wave functions at the origin on the mass spectrum using the decay constant \cite{Afonin:2007gd}. For the large radial excitation $n$,  we have 

\begin{equation}
    f_n^2\propto \frac{d M_n^2}{dn} \rightarrow \left|\Psi^{\text{rel.}}_n(0)\right|^2\propto \frac{d}{dn }E_n^3.
\end{equation}

The sum rules exposed in \cite{Afonin:2007gd} are relativistic by definition. In the relativistic case, we must correct the notion of velocity in the bounded state. Generalizations of the WKB approximation for the quark/antiquark wave function lead to the eqn.\eqref{relcorrect}. In fact, the ratio between the binding energy and the quark mass defines the system velocity scale, and the high relativistic case is when $\frac{E_n}{m_q}\gg 1$. In this situation, we have from eqn. \eqref{relcorrect}: 

\begin{equation}
     \left|\Psi^{\text{rel.}}_n(0)\right|^2\propto \frac{d}{dn }E_n^3.
\end{equation}

The sum rules and WKB approximation agree for large $n$.

In the case of heavy vector mesons, for example, the computation of the constituent quark mass differs by less than $10\%$ when comparing the relativistic with the non-relativistic Segre formulas. Since the dimensionless fraction controls the expansion $\frac{E_n}{m_q}$, the limits of small quark masses and large excitation number are equivalent. We remark that the relativistic correction in eq. \eqref{relcorrect} accounts for relativistic kinematics of a particle only, but keeping the non-relativistic structure of the wave function. Experimental data on heavy vector mesons have information on decay constants up to $n=4$. There is no information on decay properties for higher excitations. Indeed, a fully relativistic counterpart of the present approach will be necessary to discuss the highly excited states, which represent an important consideration in future work.

\section{Conclusions}\label{concl}
In this work, we unveil the constituent quark model underlying the holographic modeling of the electromagnetic decay constants of vector mesons and how this provides a new perspective on the holographic description of hadrons in bottom-up models. In the context of heavy vector mesons, we demonstrate that the Segre formula enables the determination of constituent quark masses from the holographic meson spectrum. Therefore, we provide a holographic calculation of the quark masses. 

Decay constants play a crucial role. Experimental data on decay constants, when used as a benchmark to fit parameters in holographic models, imply that we are taking the $e^+e^-$ decay width as an input in the model. We used the mass spectrum and $e^+e^-$ decay width to define the dilaton parameters in the holographic model, as was done in ref. \cite{MartinContreras:2021bis}.  We obtained the constituent quark masses and the three-photon decay widths for the ground and excited states of heavy quarkonia. We also used the ground state three-gluon decay width to define $\alpha_s$, calculating the three-gluon and one-photon-two-gluon decay widths for the ground and excited states. Calculations of quark masses and strong coupling constants in a holographic model are crucial. However, it also leads to predictions for decay widths associated with the three vector meson channels discussed above: $ggg$,   $\gamma\gamma\gamma$, and  $gg\gamma$.  In this sense, AdS/QCD models exhibit greater predictability than previously explored, as the phenomenological landscape for these models has been extended to include vector meson decay widths, the strong coupling constant, and constituent quark masses.  

The status of experimental data on decay widths is summarized in Table \ref{tab:one}, where we can observe that there is available data for most of the observables we discuss, except for bottomonium three-photon decay, for which there is no data available. The results we find using this bottom-up model are not accurate. However, they are consistent because our predictions are in the same order of magnitude, exhibiting the expected phenomenological behavior. This opens the possibility of further holographic upgrades. 

In the constituent quark model, the decay width  ratio of different radial excitations is channel-independent when expressed as a function of the meson spectrum:

\begin{equation}   
 \frac{\Gamma_{V_{n_1}}}{\Gamma_{V_{n_2}}}=  
   \frac{f_{n_1}^2\,M_{n_1}}{f_{n_2}^2\,M_{n_2}}. 
\end{equation}

The mass increases with the radial excitation number, while the experimental measurements of decay widths for the $e^+e^-$, $\gamma gg$, and $ggg$ channels decrease, as shown in Table \ref{tab:one}. Thus, decreasing the decay constant with $n$ is a key to the holographic modeling of mesons. Our results show a decreasing with $n$ in the decay widths because they capture the correct phenomenological behavior expected for the decay constants. A holographic model that does not capture the decreasing behavior of decay constants will lead, as a consequence, to an inconsistent behavior of all the decay widths. It also reveals a limitation of such an approach. The fraction of branching ratios for different radial excitations depends on the meson family only, and it is independent of the decay channel. However, the experimental data for charmonium and bottomonium are sufficient to conclude that these fractions are, in fact, channel-dependent. Thus, in this sense, even if we choose parameters fitting precisely one decay channel, the same precision will not be shared with the other channels. In summary, using non-relativistic quantum mechanics to describe meson decay enables us to obtain theoretical results for a set of new observables. However, it will not be sufficient to provide a fine-tuned description of them.

\begin{acknowledgments}
 M. A. Martin Contreras would like to acknowledge the financial support provided by the National Natural Science Foundation of China (NSFC) under grant No. 12350410371. A. V is partially supported by Centro de Física Teórica de Valparaíso (CEFITEV) and by FONDECYT (Chile) under Grant No. 1251106. Saulo Diles thanks to the Conselho Nacional de Desenvolvimento Científico e Tecnológico (CNPq), Brazil, Grant No. 406875/2023-5.

\end{acknowledgments}
\bibliography{apssamp}.
\end{document}